\documentclass[]{IEEEtran} 
\usepackage{amsmath,amssymb}
\usepackage{mathtools}
\usepackage{graphicx,epsfig}
\usepackage{setspace} 
\usepackage{multirow}
\usepackage{mwe}
\usepackage{algpseudocode}
\usepackage[ruled]{algorithm2e}

\usepackage{verbatim} 

\SetKwInput{KwInput}{Input}               
\SetKwInput{KwOutput}{Output}
\begin{document}
\pagestyle{empty}    
\title{Factory Automation: Resource Allocation of an Elevated LiDAR System with URLLC Requirements}
\author{
\IEEEauthorblockN{Nalin~Jayaweera, Dileepa~Marasinghe,~Nandana~Rajatheva,~
        and~Matti~Latva-aho.~} \\
        \IEEEauthorblockA{\textit{Centre for Wireless Communications,} 
\textit{Univeristy of Oulu, Finland}\\
E-mail: \{nalin.jayaweera, dileepa.marasinghe, nandana.rajatheva, matti.latva-aho\}@oulu.fi}
}

\maketitle
\thispagestyle{empty}
\begin{abstract}
Ultra-reliable and low-latency communications (URLLC) play a vital role in factory automation. To share the situational awareness data collected from the infrastructure as raw or processed data, the system should guarantee the URLLC capability since this is a safety-critical application. In this work, the resource allocation problem for an infrastructure-based communication architecture (Elevated LiDAR system/ ELiD) has been considered which can support the autonomous driving in a factory floor. The decoder error probability and the number of channel uses parameterize the reliability and the  latency in the considered optimization problems. A  maximum decoder error probability minimization problem and a total energy minimization problem have been considered in this work to analytically evaluate the performance of the ELiD system under different vehicle densities.     
\end{abstract}

\begin{IEEEkeywords}
Elevated LiDAR, V2X, Factory Automation, Resource Allocation, URLLC, 5G.
\end{IEEEkeywords}

\section{Introduction} 
In the past, robots in the factory floors were navigated by the aid of magnetic strips. The development of advanced sensing technologies and the advent of artificial intelligence enabled standalone robots with the vision aided navigation. The navigation decisions are based on processing their observations captured, when scanning the environment with the help of a set of on-board sensors as in modern Autonomous Vehicles (AVs)\cite{greimel_2017}.Sensing devices like radars, LiDARs and cameras simultaneously capture situation awareness data which are used to generate a floor plan around the vehicle to navigate without any  collision. Recently, wireless communication became a key enabler for factory automation to share information among moving vehicles\cite{holfeld}. The vehicle to everything(V2X) communication provides the side information required to minimize blind spots. However, sharing the huge amount of sensor data generated by the sensors imposes a huge burden on communication links. In contrast to the sensing at the vehicle, infrastructure-based sensing and processing is a better approach to automate a well-controlled environment like a factory environment. In our previous work, we proposed an infrastructure-based sensing and communication architecture to facilitate autonomous driving \cite{mybase}. In this paper, we concentrate on one of the potential use case of the proposed system.

Until now, all generations of cellular systems (2G, 3G, 4G) focused to improve the data rates significantly compared to the previous generation. As the next-generation cellular system, 5G and beyond communication systems play a wider role compared to the previous generations. Conventionally, these networks focus on broadband services with higher data rates. However, URLLC and machine type communications are also two main services focused by the 5G and beyond networks. Generally, URLLC services require reliability higher than $10^{-5}$ with an end-to-end latency less than 1 ms \cite{Durisi}. Further, mission-critical applications like factory automation may require reliability tighter than $10^{-9}$ in terms of decoder error probability \cite{based}. These facts reflect that industrial automation also requires extremely high reliability and low latency as in AVs in the road infrastructure.

\subsection{Infrastructure based communication architecture to facilitate autonomous driving.}

\begin{figure}[ht]
\centerline{\includegraphics[scale=0.27,bb=850 0 0 450]{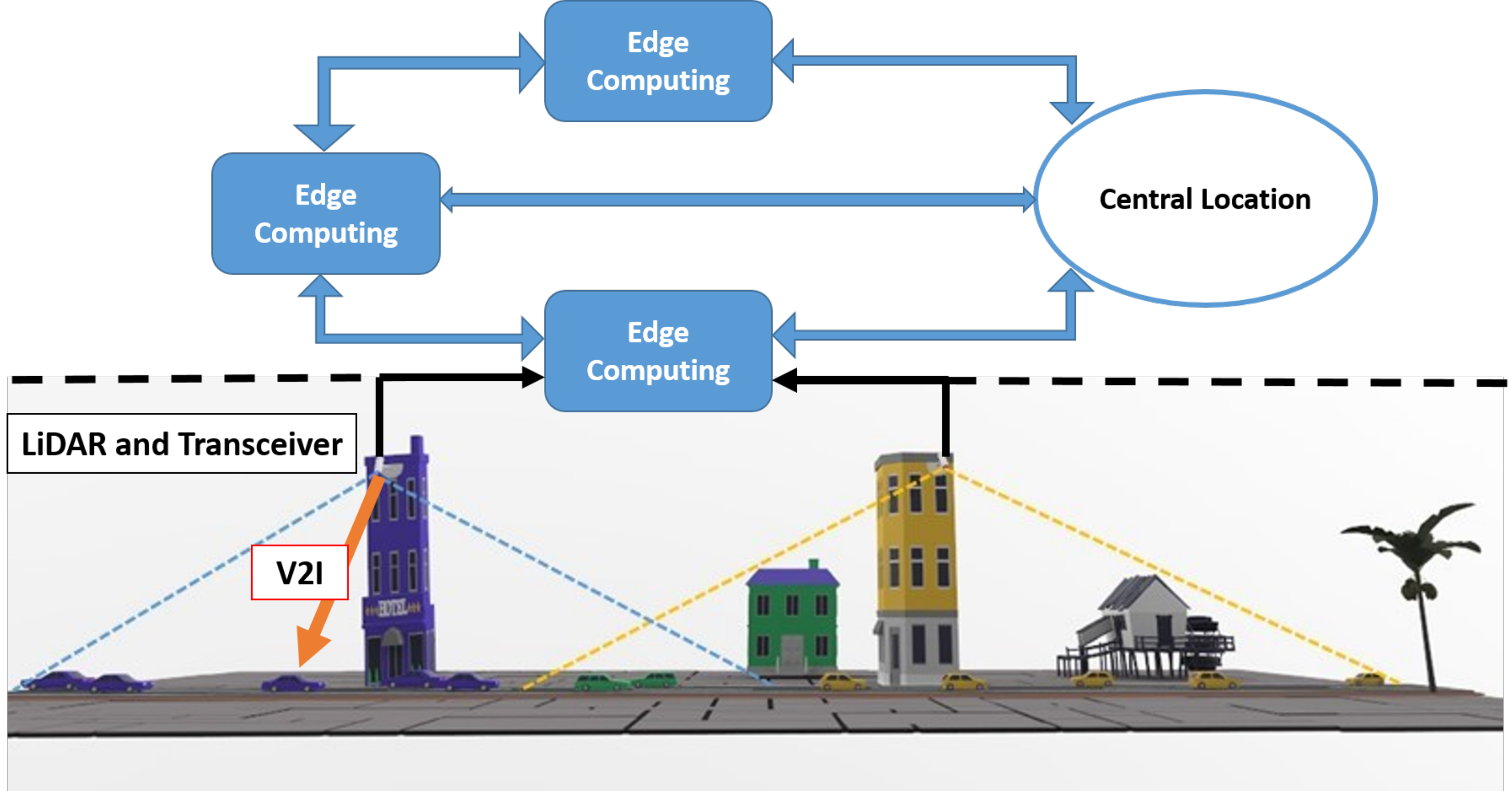}}
\caption{The proposed ELiD system architecture.}
\label{fig:architecture}
\end{figure}

The existing AV architecture has many drawbacks such as heavy in-vehicle signal processing and the burden of storing data. Collecting the situational awareness data from the infrastructure will minimize these drawbacks. Further, it can also minimize the V2X communication requirements. In our previous work we proposed a system consisting of a coordinated set of LiDARs mounted in elevated positions (ELiDs) to capture situational awareness data from a bird's-eye view \cite{mybase} (Fig.\ref{fig:architecture}). In the architecture, each LiDAR generates a 3D point cloud of the designated road sections and a central location (CL) is responsible in collecting the high-resolution point clouds (through high speed backhaul connections) and fuse them together to generate a global real-time map. Object tracking and path planning are done using the generated map in the CL. The control information needed for the vehicle is  sent back to the corresponding ELiD modules and the ELiDs transmits control information in small packets to the vehicles ensuring required latency and reliability.   

\subsection{Low Latency with Short Packet Transmissions} 
This work mainly focuses in delivering the control information packets from ELiDs to vehicles with required reliability and latency. Multiple delay sources ultimately add up to the total latency ($L$) of the downlink communication link \cite{pocovi16} as

\begin{equation}
    L =  d_{Q} + d_{ELiD} + d_{FA} + d_{Tx} + d_{v}
\end{equation}
where,  $d_{ELiD}$, $d_{v}$ are processing delays of ELiD and vehicle, $d_{Q}$, $d_{FA}$ are the queuing delay and frame alignment delay and $d_{Tx}$ represents the transmission delay or transmission time interval (TTI) needed to transmit the packet. Both $d_{ELiD}$ and  $d_{v}$ can be considered as constants from the communication point of view. Even though air interface latency is only a single latency component among all, reduction of it is mandatory to minimize the total latency ensuring URLLC. Use of short packets for the transmission is one key idea to enable URLLC. Therefore, a small packet size such as 20 bytes is used in URLLC services. Short packets can reduce the above-mentioned delay components by a significant margin, thus making efficient use of the resources ensuring latency requirements. Studies on short packets showed that certain adjustments to the classical information theoretic principles is needed to ensure reliability.

\begin{figure}[ht]
\centerline{\includegraphics[scale=0.35,bb=500 0 0 350]{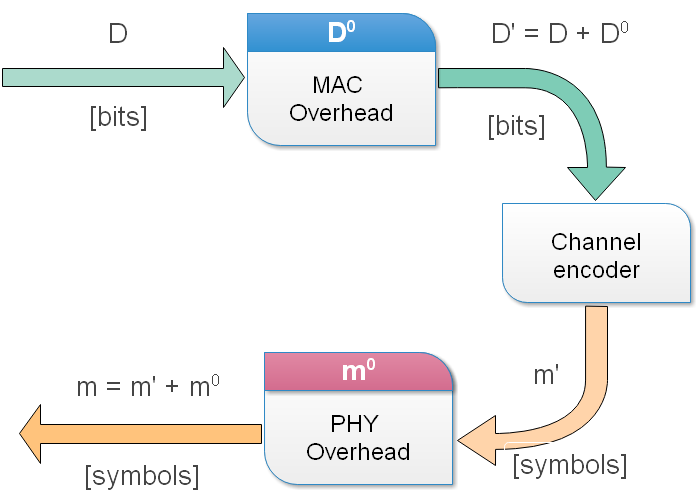}}
\caption{Packet generation as a block diagram.}
\label{fig:packetGenerartion}
\end{figure}

The mapping between information payload and transmitted signals over the channel is defined as the channel code. The responsibility of the receiver is to recover the transmitted information with low probability of error using the distorted received signal. Information theory states that, as the packet length (or the number of channel uses required to transmit the information payload) tends to a large number, there exists a channel code which can reconstruct original information at the receiver with a small probability of error \cite{Shannon}. The Fig. \ref{fig:packetGenerartion} elaborates the conversion of information bits to the transmitted symbols at the transmitter \cite{Durisi}. 

The Fig. \ref{fig:packetGenerartion} shows the translation of $D$ information bits that should be sent to the receiver. $D^0$ bits are added by the medium access control (MAC) protocol to the original information bits to make a total of $D^{'}$ bits. After that the channel encoder translates  $D^{'}$ bits to $m^{'}$ symbols. Physical layer adds  $m^0$ more symbols as its overhead which results in a total of $m$ symbols or channel uses to transmit $D$ information bits. 

In most of the communication systems $D^{0} << D$ and $m^{0} << m^{'}$ which we also assume in this work. The ratio
\begin{equation}
    R =  \frac{D}{m}
    \label{eqn:myrate}
\end{equation}
is the rate which is the number of information bits per complex symbol with a dimension of bits per second per bandwidth  \cite{Durisi}. The famous Shannon's capacity equation gives the maximum rate that the transmitter and receiver pair can achieve,
\begin{equation}
    R =  \log_2\big(1+\gamma \big)
    \label{eqn:rate}
\end{equation}
where $\gamma$ is the SNR at the receiver. As the block length is restricted to a finite number in the short packet transmission, the possibility of making an error in the decoder is no longer negligible. So, the maximum achievable rate for short block lengths becomes a function of the decoder error probability ($\epsilon$) and the short block length ($m$). Based on the Polyanskiy's approximation for the short block length \cite{Polyanskiy}, the maximum coding rate can be approximated as
\begin{equation}
R(m, \epsilon) \approx \Bigg[\log_2(1+\gamma)-\sqrt\frac{v}{m}\frac{Q^{-1}(\epsilon)}{ln(2)}\Bigg]
\label{eqn:polyR}
\end{equation}
where $v$ is channel dispersion, which is a function of SNR.
\begin{equation}
v = 1 - \frac{1}{(1 + \gamma )^2}
\label{eqn:cdisp}
\end{equation}
$Q^{-1}(\epsilon)$ is the inverse of the Q-function. The second term in the right-hand side of (\ref{eqn:polyR}) is a penalty given for the short block length. Clearly, as $m$ goes to infinity, (\ref{eqn:polyR}) reduces to (\ref{eqn:rate}).

\section{System Model and Problem Formulation}

This section presents the system model and problem formulation. We consider a factory floor environment where multiple AVs perform their responsible tasks while navigating inside the factory. As we discussed earlier, the ELiD module is mounted with an elevation with bird's eye view. In this system, we consider a single ELiD module in the ELiD system to facilitate the autonomous navigation in the factory floor. Even though communication is bidirectional, we focus only on the downlink communication in this problem.     

In the system, the ELiD module is responsible for establishing URLLC links with the set of AVs $\mathbf{V}$ where $|\mathbf{V}|$ is equal to $n$. The ELiD communicates periodically with vehicles to steer them safely to the required destinations. All the vehicles are treated equally, and they require similar type of information (steering angle, acceleration) for navigation. The CL generates the required information and sends to the ELiD for the transmission. The ELiD transmits the information as small data packets where the transmission should be completed within time $t_{max}$ to satisfy URLLC conditions. Symbol time ($t_{sym}$) decides the total number of symbols ($M = t_{max}/t_{sym}$) that can be transmitted within the transmission time. The system bandwidth (B) of such a system can be found as reciprocal of $t_{sym}$

\begin{figure}[h]
\centerline{\includegraphics[scale=0.32,bb=850 0 0 350]{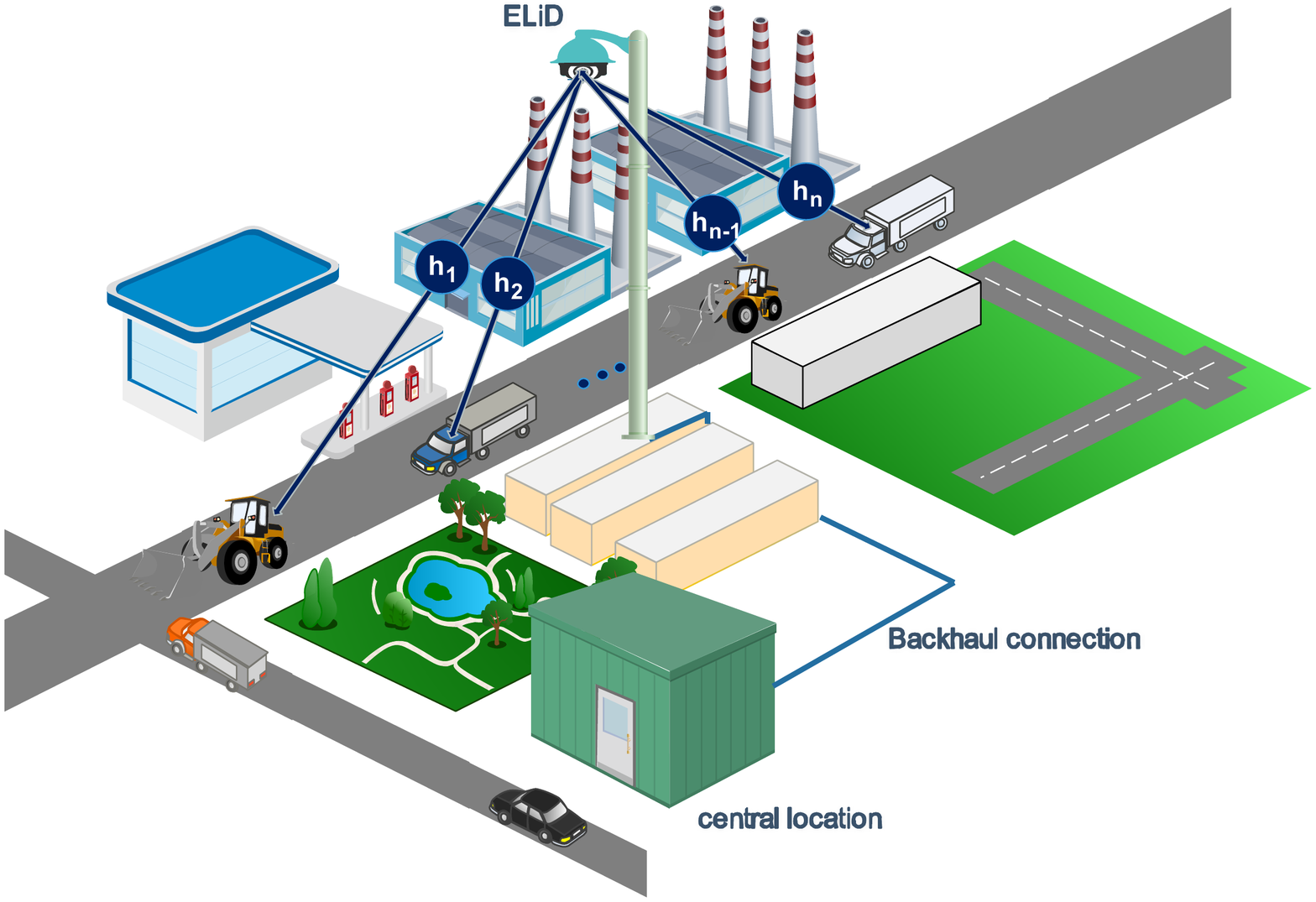}}
\caption{ELiD system facilitated autonomous navigation on a factory floor.}
\label{fig:facfloor}
\end{figure}

Let us consider that all the vehicles require a packet with $D$ bits periodically for navigation purposes. All bits should be transmitted within $M$ symbols and all the vehicles should be served within $M$ symbols. If $D$ bits required for the $i^{th}$ vehicle ($i \in V$) are distributed among $m_i$ symbols, approximated rate for the $i^{th}$ vehicle at the ELiD transmitter (channel state information at transmitter is assumed to be known) can be expressed by combining (\ref{eqn:myrate}) and (\ref{eqn:polyR}) as \cite{based}

\begin{equation}
\frac{D}{m_i} = \Bigg[\log_2(1+\gamma_i)-\sqrt\frac{v}{m_i}\frac{Q^{-1}(\epsilon_i)}{ln(2)}\Bigg]
\label{eqn:ratei}
\end{equation}
where $\gamma_i$ and $\epsilon_i$ represent approximated SNR and decoder error probability at the receiver of the $i^{th}$ vehicle. To make communication reliable, SNR should be sufficiently high. According to (\ref{eqn:cdisp}), we can claim that $v$ tends to 1 in high SNR regime. We consider an orthogonal multiple access technique to mitigate interference. The received signal at the $i^{th}$ vehicle can be expressed as
\begin{equation}
y_i = \sqrt{p_i}h_ix_i + w_i
\label{eqn:channel}
\end{equation}
where $p_i$ is the allocated power, $h_i$ is the channel and $x_i$ is the transmitted signal to the $i^{th}$ vehicle. $w_i$ is zero mean additive white Gaussian noise (AWGN) with variance $\sigma_i^2$. A graphical view of the system model is shown in the Fig.\ref{fig:facfloor}.

The approximated SNR at the receiver of the $i^{th}$ vehicle is
\begin{equation}
\gamma_i = \frac{p_i |h_i|^2}{\sigma_i^2}
\label{eqn:SNR}
\end{equation}
Using (\ref{eqn:ratei}) and (\ref{eqn:SNR}), 
\begin{equation}
\begin{aligned}
\epsilon_i & = Q\Bigg[ln(2)\sqrt{m_i}\Big(\log_2(1+\frac{p_i|h_i|^2}{\sigma_i^2})-\frac{D}{m_i}\Big)\Bigg]
\end{aligned}
\end{equation}
This shows that the decoder error probability is a function of SNR, block length and packet size. Since there are $n$ vehicles in the system, let us define $\boldsymbol{\epsilon} = [\epsilon_1, \epsilon_2,., \epsilon_n]^T$,$ \boldsymbol{\sigma} = [\sigma_1, \sigma_2,., \sigma_n]^T$, $ \boldsymbol{P} = [p_1, p_2.., p_n]^T$ and $\boldsymbol{M} = [m_1, m_2,., m_n]^T$. 

\subsection{Maximum decoder error probability minimization problem}

In order to guarantee system reliability, all $n$ vehicles in the system should have a minimum decoder error probability. The objective function of the resource allocation problem can be expressed as
\begin{equation}
\begin{aligned}
& \min_{\forall p_i,m_i} \Big(\max_{\forall i \in V} \epsilon_i\Big)\\
\label{eqn:obj1}
\end{aligned}
\end{equation}

Since $Q$-function is a decreasing function, (\ref{eqn:obj1}) can be reformulated as

\begin{equation}
\begin{aligned}
& \max_{\forall p_i,m_i} \Big(\min_{\forall i \in V} g(\gamma_i,m_i,D)\Big)\\
\end{aligned}
\end{equation}

This formulation will be able to maximize the reliability of the system by minimizing the maximum decoder error probability of the vehicle experiencing it. Consider the infimum of $g(\gamma_i,m_i,D) $ $\forall i \in V$ as $-s$. An optimization problem can be formulated as
\begin{subequations}
\begin{alignat}{2}
& \min_{\boldsymbol{P}, \boldsymbol{M}, s} s\\
\textrm{s.t :}
&\quad ln(2)\frac{D}{\sqrt{m_i}} - ln(1+\frac{p_i|h_i|^2}{\sigma_i^2})\sqrt{m_i} - s \leq 0, \quad \forall i \in V \label{eqn:c1} \\
&\quad  \left\lVert\boldsymbol{M}\right\rVert_1 \leq M \label{eqn:c2}\\
&\quad \boldsymbol{P}^T \boldsymbol{M} \leq E_{tot} \label{eqn:c3}\\
&\quad m_i \in \mathbb{Z^+} \quad \forall i \in V \label{eqn:c4}
\end{alignat}
\end{subequations}

In the above formulation, constraint (\ref{eqn:c1}) should be satisfied to meet the required reliability of communication for all vehicles. (\ref{eqn:c2}) represents the latency constraint where the sum of block lengths should be less or equal to the total number of channel uses available in the system. Constraint (\ref{eqn:c3}) presents the total energy constraint of the ELiD system. To reduce the search complexity and to avoid infeasible solutions, lower and upper bounds for $m_i$ are defined. The minimum $m_i$ satisfying the following equation is the lower bound ($m_i^{lb}$) of $m_i$ \cite{based}.

\begin{equation}
\begin{aligned}
 h_i E_{tot} > m_i (2^{D/m_i}-1)
 \label{eqn:lb}
\end{aligned}
\end{equation}

According to constraint (\ref{eqn:c3}) and (\ref{eqn:lb}), upper bound of $m_i$ can be calculated as
\begin{equation}
\begin{aligned}
m_i^{ub} = M - \sum\limits_{\forall j \in V/i} m_j^{lb}
 \label{eqn:ub}
\end{aligned}
\end{equation}
The feasible set for $m_i$ can be found as [$m_i^{lb}$,$m_i^{ub}$]. 

This formulation leads to less decoder error probabilities ($<< 10^{-9}$) when $E_{tot}$ is set to higher value while SNR at the receivers may vary in the range of few 10's. 

\subsection{Total energy minimization problem}

In this subsection, the problem has been reformulated such that the system will minimize the total energy consumption while achieving a targeted decoder error probability ($s_{t} = 10^{-9}$).  
\begin{subequations}
\begin{alignat}{2}
& \min_{\boldsymbol{P}, \boldsymbol{M}, s} \boldsymbol{P}^T \boldsymbol{M}\\
\textrm{s.t :}
&\quad ln(2)\frac{D}{\sqrt{m_i}} - ln(1+\frac{p_i|h_i|^2}{\sigma_i^2})\sqrt{m_i} - s \leq 0, \quad \forall i \in V \label{eqn:c21} \\
&\quad  s - s_t  \leq 0 \label{eqn:c22}\\
&\quad  \left\lVert\boldsymbol{M}\right\rVert_1 \leq M \label{eqn:c23}\\
&\quad  min\{\boldsymbol{P}\} \geq 0 \label{eqn:c24}\\
&\quad m_i \in \mathbb{Z^+} \quad \forall i \in V \label{eqn:c25}
\end{alignat}
\end{subequations}

The constraint (\ref{eqn:c22}) guarantee that the targeted decoder error probability has been achieved. The rest of the constraints hold as before. The symbol sharing algorithm has been proposed to solve the problem as shown below, 

\begin{algorithm}[h]
\DontPrintSemicolon
  
  \KwInput{$n$,\textbf{h},\textbf{$\sigma$},D,M,$s_t$,$\alpha$ }
  \KwOutput{$\boldsymbol{P}^*$,$\boldsymbol{M}^*$,$E_{tot}^*$}
  
  $m_i$ = $M/n  \quad \forall i \in V$ \\
  $E_{tot} \rightarrow \infty $\\
 \While{True}{
  minimize $\boldsymbol{P}^T$ $\boldsymbol{M}$ over $\boldsymbol{P}$ for constant $\boldsymbol{M}$ \\
  $\boldsymbol{P^*}$ = $optimal\{\boldsymbol{P}\}$\\
  $E_{tot}^* = optimal\{\boldsymbol{P}^T \boldsymbol{M}\}$\\

  \eIf{$E_{tot}^* \leq E_{tot}$}{
   		 $E_{tot} = E_{tot}^*$\\
   		 i = $User\{max\{\boldsymbol{P^*}\}\}$ \\
   		 $\boldsymbol{M}(i) = \boldsymbol{M}(i) + \alpha $\\
   		 j = $User\{min\{\boldsymbol{P^*}\}\} $\\
   		 $\boldsymbol{M}(j) = \boldsymbol{M}(j) - \alpha $\\
   		}
   		{
   		$\boldsymbol{M^*} = \boldsymbol{M}$\\
   		break\\
   		}
 }
\caption{Symbol sharing algorithm for total energy minimization of the ELiD system}
\end{algorithm}

The number of vehicles in the system, channel matrix, noise power, number of information bits to be transmitted, the total number of channel uses (symbols) available and the number of symbols shared in one iteration ($\alpha$) are the inputs to the algorithm. Outputs are optimum user-specific powers and corresponding block lengths for each user. Initially, set $\boldsymbol{M}$ to an equally allocated vector and calculate the optimum energy over variable $\boldsymbol{P}$ for a given $\boldsymbol{M}$. As the next step, $\alpha$ number of symbols transfer from the user who requires minimum power to the user who requires maximum power at that iteration if and only if the optimal energy is lesser than the previous iteration. The iterations are carried out until the total energy reaches the minimum value.

\section{Simulation Results} 
This section presents the simulation results to describe the behaviour of the system. The ELiD system can cover a straight road section of 397 m. In a factory floor environment, we assume less mobility and the number of vehicles may not exceed 10. The bandwidth of the ELiD system is set to be 1 MHz and the packet size is 20 bytes. The noise power is assumed to be -180 dBm/Hz and the pathloss model is $35.3 + 10 \times 3.76 \times \log_{10}(d)$ dB. A rician fading has been considered due to line of sight communication. 

\begin{figure}[ht]
\centerline{\includegraphics[scale=0.6]{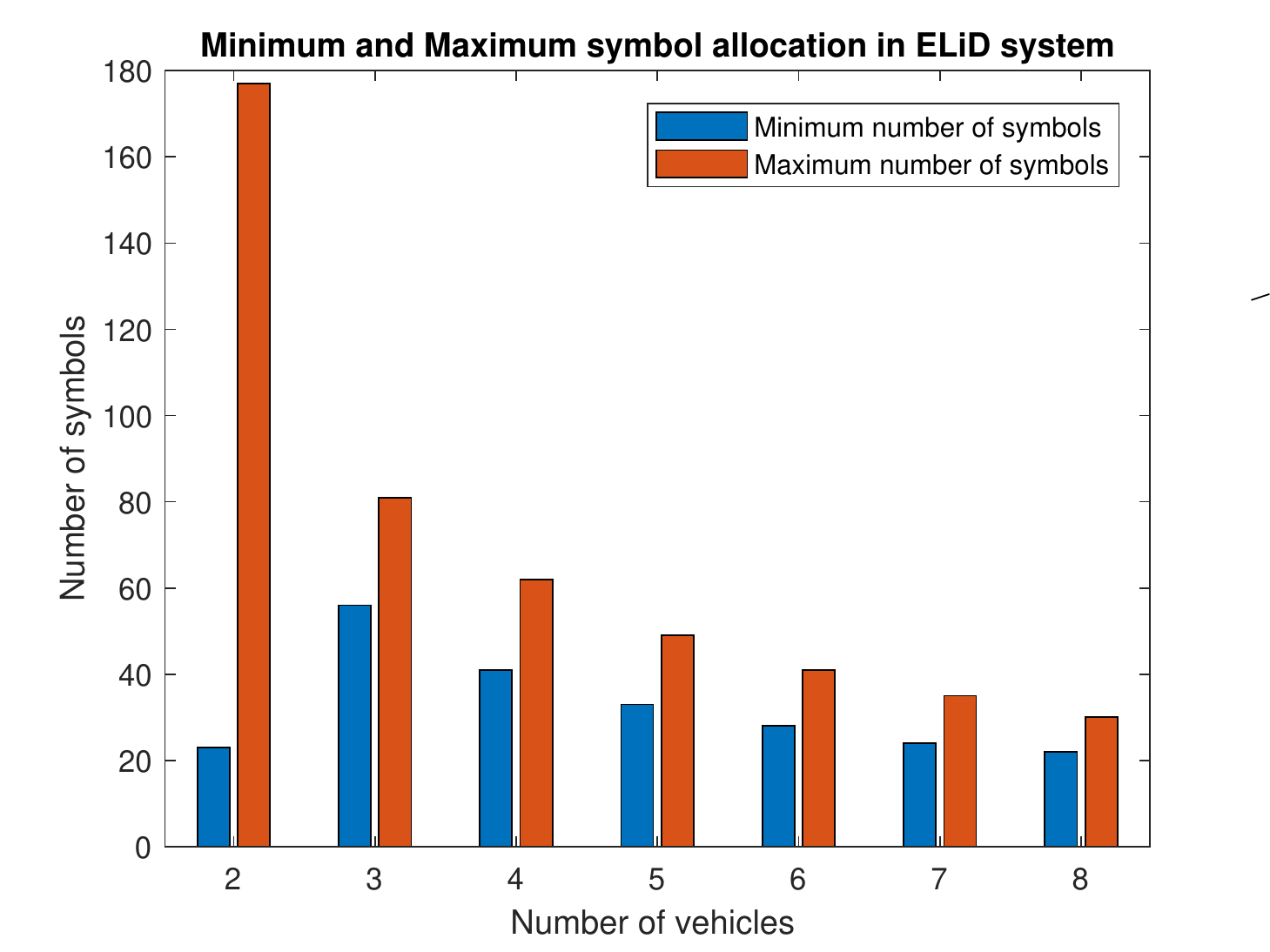}}
\caption{Variation of minimum/ maximum user-specific symbol allocations for different vehicular densities (M = 200).}
\label{fig:variationofM}
\end{figure}

The variability of the maximum and the minimum number of symbols which are allocated by the ELiD system, under constant user-specific power allocation for different vehicular densities is shown in Fig. \ref{fig:variationofM}. As the number of vehicles in the system increase, user-specific symbol allocation converges to a constant. When the number of serving vehicles is high, the optimal $m_i$ can be approximated to $M/n$ by reducing the complexity of the problem.

\begin{figure}[h]
\centerline{\includegraphics[scale=0.55]{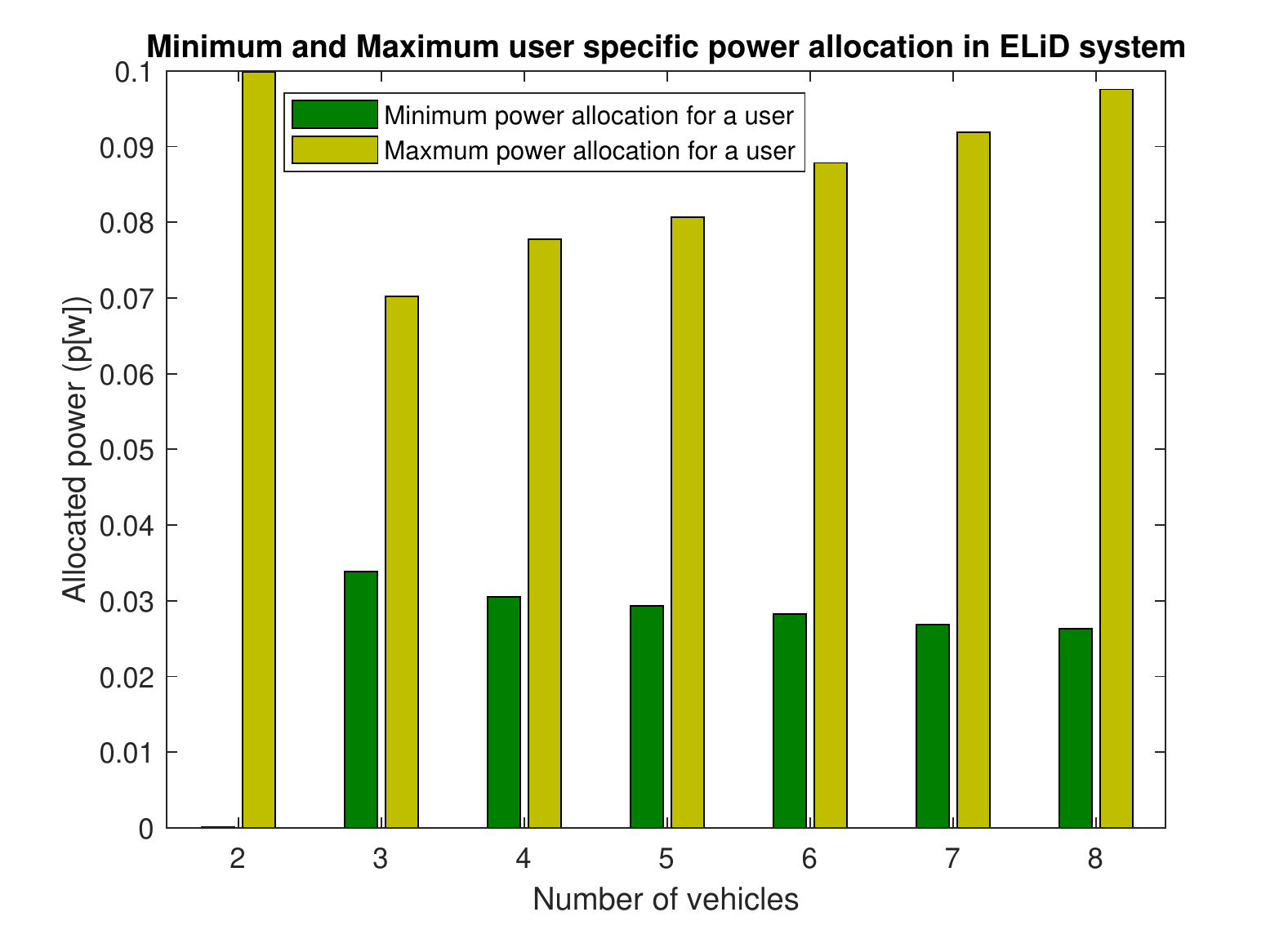}}
\caption{Variation of minimum/ maximum user-specific power allocations for different vehicular densities ($E_{tot} = 10J$).}
\label{fig:variationofP}
\end{figure}

\begin{figure}[ht]
\centerline{\includegraphics[scale=0.45]{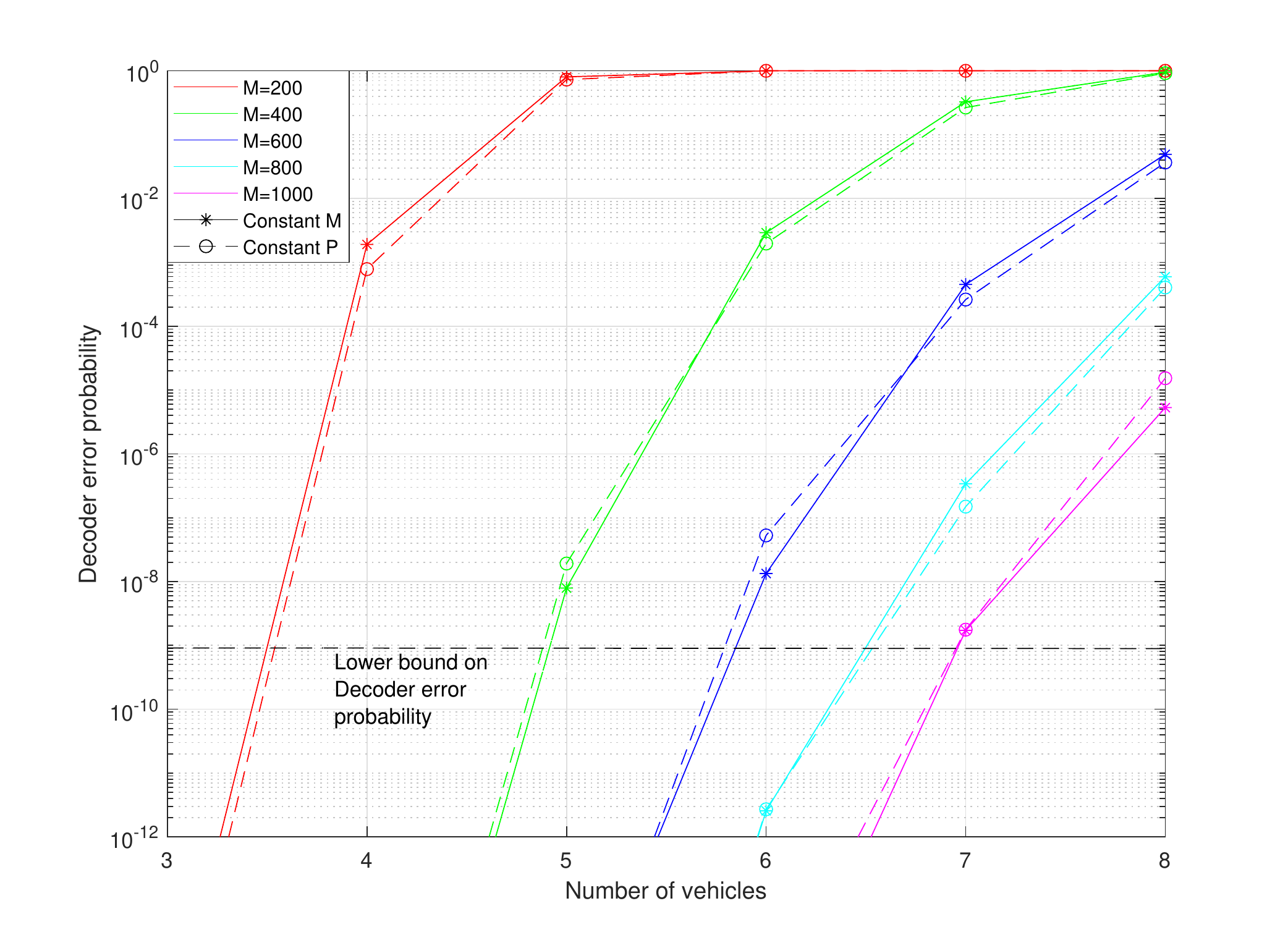}}
\caption{Maximum decoder error probability in the system under constant user-specific power allocations and constant user-specific symbol allocation for varying vehicle densities(Packet size is 20 bytes).}
\label{fig:comp}
\end{figure}

Fig. \ref{fig:variationofP} shows the variability of the maximum and minimum power allocations for different vehicle densities under constant symbol allocation while Fig. \ref{fig:comp} presents the variation of minimized maximum decoder error probability under constant user-specific power allocation case and constant symbol allocation case for different vehicle densities. Under both cases, minimized maximum decoder error probability is nearly equal. 

\begin{figure}[ht]
\centerline{\includegraphics[scale=0.48]{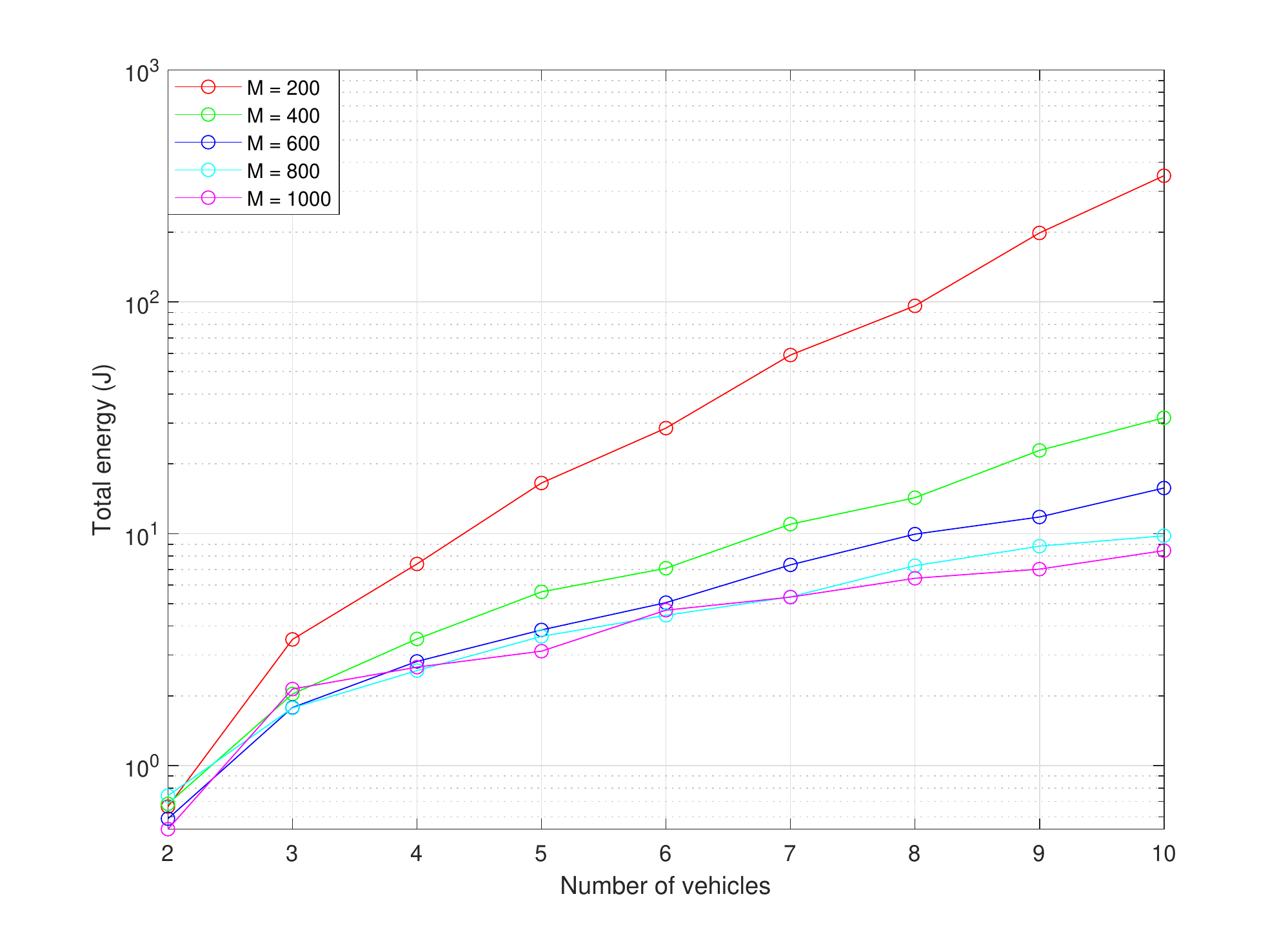}}
\caption{Variation of $E_{tot}$ against number of vehicles ($\alpha$ = 1, D = 160 bits).}
\label{fig:variationofE}
\end{figure}

\begin{figure}[ht]
\centerline{\includegraphics[scale=0.5]{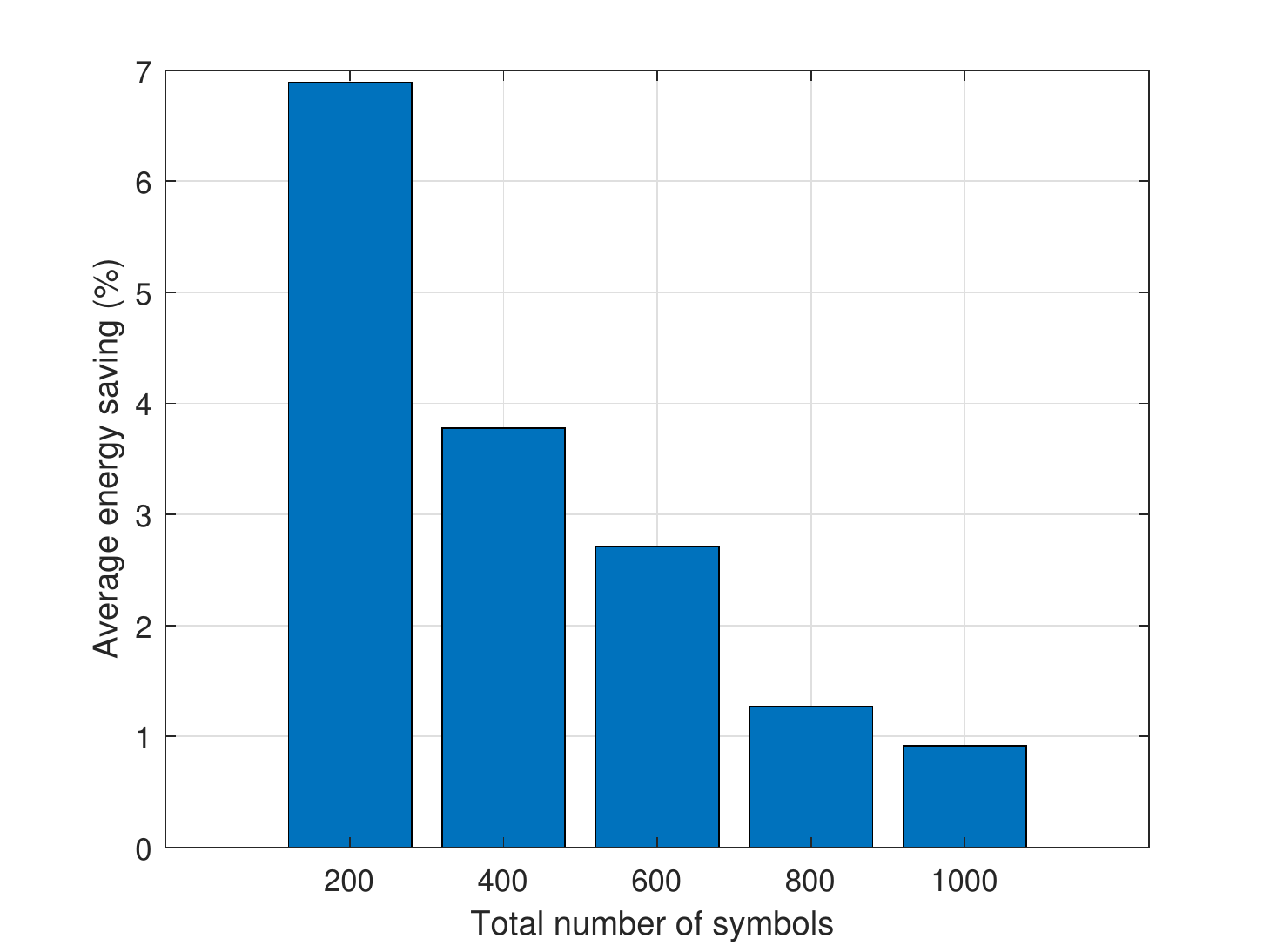}}
\caption{Average energy saved for varying symbol availability ($\alpha$ = 1, D = 160 bits).}
\label{fig:savedE}
\end{figure}

Fig. \ref{fig:variationofE} shows the variation of the minimum energy required against the number of vehicles served by the ELiD. As available symbols reduce from 1000 to 200 (Latency 1 ms to 0.2 ms), the total energy requirement of the ELiD system increases rapidly. The amount of energy saved by the proposed symbol sharing algorithm compared to the equal symbol allocation approach is presented in Fig. \ref{fig:savedE}. 
The system can handle any number of vehicles with required reliability and latency by varying total energy requirement. The number of iterations needed for the convergence of the algorithm depends on the value of $\alpha$.      

\section{Conclusion} 
This work focused on a resource allocation problem in an ELiD system which enables the autonomous driving in a factory floor setting with less mobility and less number of vehicles to be served. Short packet transmissions have been considered to achieve the required reliability and latency requirements. We formulated an optimization problem to minimize the maximum decoder error probability of the ELiD system. As the number of vehicles in the system increased, the system was not capable of delivering the required reliability as the total energy of the system is fixed. As the next step, the problem was reformulated as a total energy minimization problem and an algorithm has been proposed to decide the user-specific powers and user-specific symbol allocations depending on the latency, reliability and the number of vehicles in the system. The proposed algorithm has the ability to save a significant amount of energy compared to the equal symbol allocation approach (Fig. \ref{fig:savedE}).

\bibliographystyle{IEEEtran}  
\bibliography{Sec_6_Bibliography.bib}

\end{document}